\documentclass{article}

\PassOptionsToPackage{numbers, compress}{natbib}
\usepackage{neurips_2025}

\usepackage{amsmath,amsfonts,bm}

\def\eqref#1{equation~\ref{#1}}
\def\1{\bm{1}}

\def\vh{{\bm{h}}}

\DeclareMathAlphabet{\mathsfit}{\encodingdefault}{\sfdefault}{m}{sl}
\SetMathAlphabet{\mathsfit}{bold}{\encodingdefault}{\sfdefault}{bx}{n}

\usepackage{microtype}
\usepackage{graphicx}
\usepackage{tikz}
\usetikzlibrary{arrows.meta, positioning, shapes.geometric, backgrounds}
\usepackage{subfigure}
\usepackage{booktabs} % for professional tables
\usepackage{amsmath,amssymb,amsfonts}
\usepackage{mathtools}
\usepackage{enumerate}
\usepackage{nicefrac}
\usepackage{siunitx}
\usepackage{url}
\usepackage{hyperref}
\usepackage{csquotes}
\usepackage{multirow}
\usepackage{makecell}

\title{Electrocardiogram Classification with Transformers Using
Koopman and Wavelet Features}

\author{Sucheta Ghosh \textsuperscript{1,2}, Zahra Monfared\textsuperscript{1,2,3} \\ \\
  \{sucheta.ghosh,zahra.monfared\}@iwr.uni-heidelberg.de\\
\textsuperscript{1}Interdisciplinary Center for Scientific Computing, Heidelberg University, Germany
\\ \textsuperscript{2} Department of Mathematics and Computer Science, Heidelberg University, Heidelberg, Germany\\  
\textsuperscript{3}Department of Theoretical Neuroscience, Central Institute of Mental Health, \\Medical Faculty Mannheim,
Heidelberg University, Mannheim, Germany\\  
}

\begin{document}
\maketitle

% -------- Abstract (keep it ~4--6 sentences) --------
\begin{abstract}
Electrocardiogram (ECG) analysis is vital for detecting cardiac abnormalities, yet robust automated classification is challenging due to the complexity and variability of physiological signals. In this work, we investigate transformer-based ECG classification using features derived from the Koopman operator and wavelet transforms. Two tasks are studied: (1) binary classification (Normal vs. Non-normal), and (2) four-class classification (Normal, Atrial Fibrillation, Ventricular Arrhythmia, Block). We use Extended Dynamic Mode Decomposition (EDMD) to approximate the Koopman operator. Our results show that wavelet features excel in binary classification, while Koopman features, when paired with transformers, achieve superior performance in the four-class setting. A simple hybrid of Koopman and wavelet features does not improve accuracy. However, selecting an appropriate EDMD dictionary -- specifically a radial basis function dictionary with tuned parameters -- yields significant gains, surpassing the wavelet-only baseline and the hybrid wavelet-Koopman system. We also present a Koopman-based reconstruction analysis for interpretable insights into the learned dynamics and compare against a recurrent neural network baseline. Overall, our findings demonstrate the effectiveness of Koopman-based feature learning with transformers and highlight promising directions for integrating dynamical systems theory into time-series classification.
\end{abstract}

% -------- Page limit notes --------
% NeurIPS workshop short papers are typically limited to up to 4 pages excluding references and appendices.
% Keep everything through Section 6 within 4 pages. References and appendices may exceed the limit.
%%%%%%%%%%%%%%%%%%%%%%%%%%%%%%%
\section{Introduction and Related Work}
Electrocardiography (ECG) is central to diagnosing and monitoring cardiac conditions. With the growth of large-scale databases such as MIMIC-IV-ECG~\cite{gow2023mimic}, automated interpretation using machine learning has become increasingly important. Traditional methods rely on handcrafted time- and frequency-domain features, while deep learning models such as Convolutional Neural Networks (CNNs) and Recurrent Neural Networks (RNNs)~\cite{hannun2019cardiologist,kiranyaz2015real,rajpurkar2017arrhythmia,song2019eeggnn} achieve strong performance but face challenges in interpretability and in capturing physiological dynamics \cite{supratak2017deepsleepnet,wasim2025,williams2015data}.
%----
Recent advances point to two complementary directions. First, wavelet transforms capture localized time-frequency structure and transient arrhythmic events~\cite{addison2005wavelet,li1995detection}, and achieve robust results when combined with deep models~\cite{ubeyli2009combined,zhao2019ecg}. Second, the Koopman operator framework reformulates nonlinear dynamics as the evolution of observables in an infinite-dimensional function space, enabling linear analysis of nonlinear systems \cite{brunton2016discovering}. Dynamic Mode Decomposition (DMD) provides a data-driven finite-dimensional approximation of this operator but is limited to linear combinations of raw state variables. Extended Dynamic Mode Decomposition (EDMD) extends DMD by introducing richer dictionaries of basis functions, effectively lifting the data into a higher-dimensional feature space for more accurate linear approximations of dynamics~\cite{mezic2005spectral,rowley2009spectral,williams2015data}. While powerful for modeling nonlinear systems, EDMD’s performance depends critically on the choice of basis functions. Koopman methods have been successfully applied in fluid dynamics~\cite{brunton2016discovering,williams2015data}, neuroscience~\cite{brunton2017chaos}, oscillatory analysis \cite{acharya2013wavelet,subasi2007classification}, and time series modeling~\cite{mauroy2016linear,sinha2021koopman}, but remain underexplored in biomedical signal analysis.
%------------------
In parallel, deep sequence models, particularly Transformers~\cite{vaswani2017attention}, have become state-of-the-art for ECG~\cite{hu2022transformer,polat2007classification,wu2021transformer}, non-stationary signals \cite{zerveas2021transformer} and Electroencephalogram (EEG)~\cite{song2022transformer} classification, outperforming RNNs in modeling long-range dependencies. Hybrid approaches integrate handcrafted or spectral features with neural networks for robustness and interpretability, such as wavelets with CNNs~\cite{kiranyaz2016real}, or spectral embeddings with Transformers~\cite{zerveas2021transformer} and other nonlinear features \cite{acharya2013wavelet}. However, to the best of our knowledge, combining Koopman and wavelet features within Transformer models has not yet been explored.
%--------------------

In this work, we bridge these directions by systematically evaluating Koopman-based features with Transformers for ECG classification. Using MIMIC-IV-ECG, we study two tasks: (i) binary classification (Normal vs.\ Nonnormal), and (ii) four-class classification (Normal, Atrial Fibrillation, Ventricular Non-normalities, Conduction Block). Our experiments compare: (1) Wavelet + Transformer, (2) Koopman + Transformer, (3) Hybrid Wavelet + Koopman + Transformer, and (4) Koopman + Transformer with refined feature extraction. Results indicate that wavelet features are most effective in binary tasks, Koopman features outperform in multi-class settings, and principled improvements to Koopman features yield the strongest overall performance. Our Contributions are: (1) 
A systematic study of Koopman-based features with Transformers for ECG classification; (2) Benchmarking against wavelet baselines, clarifying their strengths; (3) Proposing improvements to Koopman feature extraction that significantly enhance classification, motivating future research on fusion and \textit{interpretability}.

%%%%%%%%%%%%%%%%%%%%%%%
\section{Method}
\begin{figure*}[ht]
\centering
\includegraphics[width=1\linewidth]{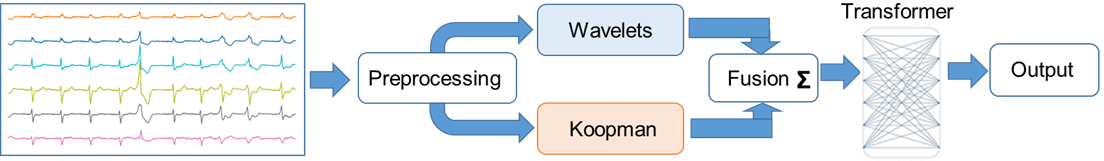}
\caption{Overall Hybrid Pipeline.}\label{fig:pip}
\end{figure*}

% Here, we investigate the use of Koopman-based feature representations, derived via Extended Dynamic Mode Decomposition (EDMD), in combination with a Transformer architecture for ECG classification. Our dataset is drawn from the MIMIC-IV-ECG database, where we utilize raw ECG waveforms and their associated diagnostic labels. \emph{We define two classification tasks of interest:} 1. Binary classification: distinguishing between normal and non-normal rhythms. 2. Four-class classification: categorizing ECG signals into Normal, Atrial Fibrillation (AFib), Ventricular arrhythmias, and Conduction Block. The \textit{experimental pipeline} follows a consistent structure across all variants (Fig.~\ref{fig:pip}). In the hybrid system, wavelet and Koopman features are concatenated before being passed to the Transformer encoder. Our pipline is as follows:
%----
We investigate the use of Koopman-based feature representations, derived via EDMD, in combination with a Transformer architecture for ECG classification. Our dataset is drawn from the MIMIC-IV-ECG database, using raw ECG waveforms and their associated diagnostic labels. \emph{We define two classification tasks:} (i) binary classification, and (ii) four-class classification; see Sect. \ref{sec:data} for details. The experimental pipeline is consistent across all variants (Fig.~\ref{fig:pip}). In the hybrid system, wavelet and Koopman features are concatenated before being passed to the Transformer encoder.  
%---------------------
%-----------------
%: ECG $\to$ Preprocessing $\to$ Feature extraction (Wavelet or Koopman EDMD) $\to$ Transformer Encoder $\to$ Classification head $\to$ Predicted labels.
% \textbf{Wavelet Feature Extraction.}
% For comparison, we construct wavelet-based features using discrete wavelet transforms (DWT). These coefficients provide multiresolution time–frequency representations of ECG signals, serving as an alternative input modality for the Transformer.

\textbf{Wavelet Feature Extraction.}
For comparison, we construct wavelet-based features using the discrete wavelet transform. The resulting coefficients provide a multiresolution time–frequency representation of ECG signals and serve as an alternative input modality for the Transformer.  
%-------------------
% \textbf{Koopman Feature Extraction.}
% We employ EDMD with radial basis function (RBF) observables to approximate the Koopman operator and derive a set of eigenfunctions and features that encode the underlying temporal dynamics of ECG signals. We are utilizing both real and imaginary part along with the magnitude and growth. These features capture nonlinear dynamics in a linearized latent space, which are then supplied as input tokens to the Transformer model. 

\textbf{Koopman Feature Extraction.}
We employ EDMD with Radial Basis Function (RBF) dictionary to approximate the Koopman operator and derive eigenspectrum that encode the temporal dynamics of ECG signals. Both the \textit{real} and \textit{imaginary components}, along with \textit{magnitude} and \textit{growth rates}, are used as features. These are then supplied as input tokens to the Transformer \cite{vaswani2017attention}.  

% In addition to feature extraction, the Koopman framework enables approximate reconstruction of the ECG signal from a subset of dominant modes. This reconstruction step serves two purposes: (1) it validates that the Koopman operator captures the essential dynamics of the waveform, and (2) it highlights interpretable dynamical patterns (e.g., oscillatory modes linked to cardiac rhythms). We report reconstruction errors alongside classification results as a measure of representational adequacy.
%------------------
\textbf{Reconstruction.} Beyond feature extraction, the Koopman framework enables approximate reconstruction of ECG signals from dominant modes. This step serves two purposes: (1) validating that the Koopman operator captures the essential dynamics of the waveform, and (2) providing interpretable dynamical patterns (e.g., oscillatory modes associated with cardiac rhythms). We report reconstruction errors alongside classification results as a measure of representational adequacy.  
%--------------------
% \textbf{Transformer-based Classifier.}
% In all cases, the feature tokens (from Koopman or wavelet transforms) are projected into a latent embedding space and passed through a Transformer encoder. The encoder outputs are aggregated and mapped to class predictions through a final classification head.

\textbf{Transformer-based Classifier.}
In all cases, the feature tokens (from Koopman or wavelet transforms) are embedded into a latent space and passed through a Transformer encoder. The encoder outputs are aggregated and mapped to class predictions via a classification head.  

\textbf{Hybrid Feature Approach.}
Given the complementary strengths of Koopman and wavelet features, we designed a hybrid system by concatenating them prior to Transformer encoding. The goal was to combine dynamical and spectral information, but it did not improve upon standalone systems.

\textbf{Refinement of Koopman Features.}
To enhance performance, we refined the Koopman feature extraction pipeline by tuning EDMD dictionary hyperparameters, including delay embedding dimension, number of RBF centers, kernel bandwidth, and spectral truncation. An ablation study showed that these refinements improved the performance of Koopman + Transformer, particularly in both binary and four-class classification tasks.  

\textbf{Comparative Baseline: RNN.}
As an additional baseline, we trained an RNN classifier on raw ECG waveforms to benchmark the proposed approaches against a conventional sequence model.
%%%%%%%%%%%%%%%%%%%%%%%%%%%%%%%%%%
\section{Experiments}
\subsection{Dataset (MIMICS-IV-ECG and  Corresponding Notes) and Setup}\label{sec:data}
We conducted experiments using the MIMIC-IV-ECG dataset, a large, publicly available collection of ECG waveforms from the MIMIC-IV clinical database. We focused on a subset of standard 12-lead ECG signals with diagnostic labels. Each record was preprocessed by resampling to a fixed rate, z-score normalization, and 50\% overlap segmentation. To reduce dimensionality while preserving temporal dynamics, features were extracted using either wavelet decomposition or Koopman mode decomposition before being input to the transformer-based classifier.
%\textbf{Label Construction}
The MIMIC-IV-ECG dataset includes diagnostic-labeled ECG waveforms linked to additional resources. For binary classification, we used labels for Normal and Non-normal rhythms, while for four-class classification, we included Normal, Atrial Fibrillation (AFib), Ventricular Arrhythmias, and Conduction Block. For details on label generation, see Appendix~\ref{sec:labelgen}. To ensure consistency, all ECG signals were segmented into fixed-length windows, normalized, and split into training, validation, and test sets (70-15-15). All models were trained using the same optimization regime with AdamW, early stopping on validation loss, and cross-entropy loss.
%%%%%%%%%%%%%%%%%%%%%%%%5
\subsection{Results: Classification, Reconstruction and Analysis}
%--------------------
\begin{table}[!htbp] 
  \label{tab:main}
  \centering
  \caption{Primary performance on the test data split (mean\,$\pm$\,std over 5 runs).}
  \label{tab:main}
  \begin{tabular}{lcc}
    \toprule
    Method & Binary~Cl~F1 & 4-Class~Cl~F1 \\
    \midrule
    Wavelet + Transformer & 0.750 $\pm$ 0.02 & 0.700 $\pm$ 0.03 \\
    Koopman + Transformer & 0.697 $\pm$ 0.01 & \textbf{0.771} $\pm$ 0.02 \\
    Hybrid (Wavelet + Koopman) + Transformer & 0.677 $\pm$ 0.01 & 0.533 $\pm$ 0.02 \\
    Koopman + Transformer (After ablation) & \textbf{0.786} $\pm$ 0.01 & 0.764 $\pm$ 0.02 \\
    RNN (Raw ECG, baseline) & 0.782 $\pm$ 0.01 & 0.700 $\pm$ 0.02 \\
    \bottomrule
  \end{tabular}
\end{table}
%---------------------
%---------------------
\begin{figure}[!htbp]
  \centering
  \includegraphics[width=0.67\linewidth]{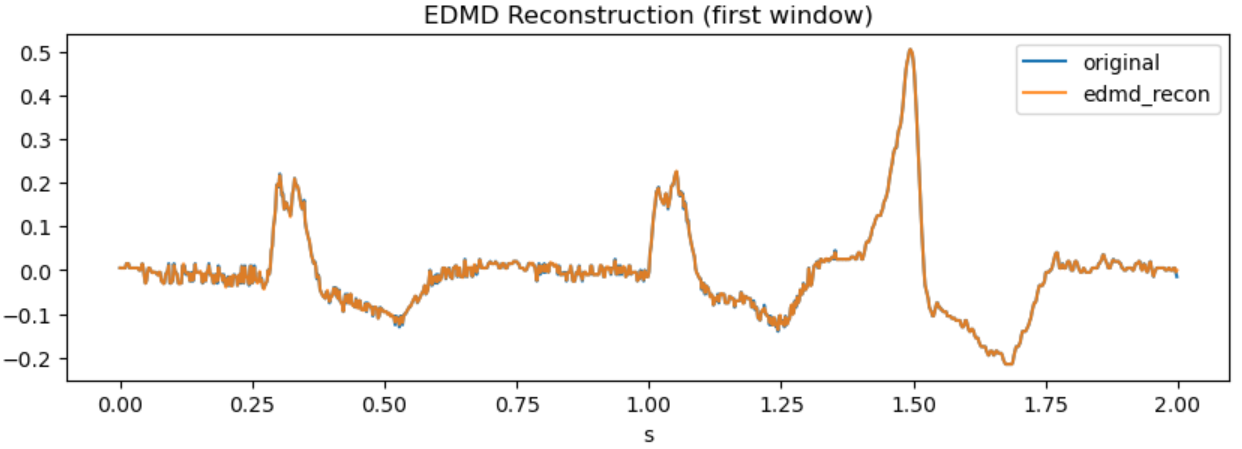}
  \caption{Reconstruction of a sample of ECG.}
  \label{fig:recon}
\end{figure}
%----------------------
%-------------------------
\begin{figure}[t]
  \centering
\includegraphics[width=0.3\linewidth]{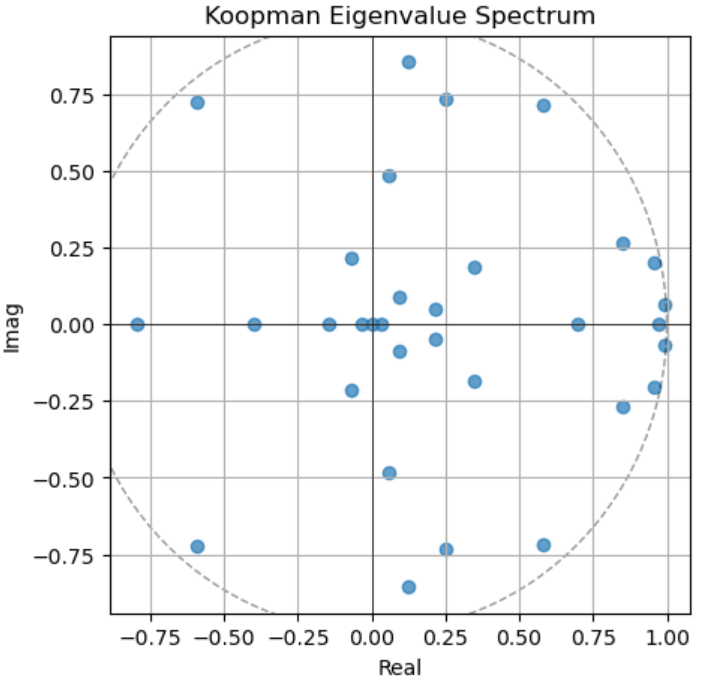}
\includegraphics[width=0.65\linewidth]{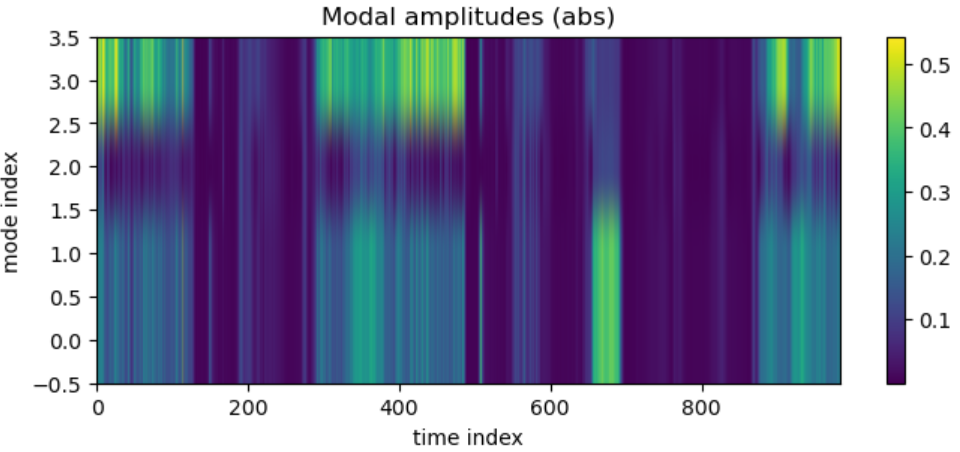}
  \caption{Left: Koopman eigenvalue spectrum of the same sample. Right: Koopman mode spectrum (modal amplitudes) of the same sample.}\label{fig:kmode}
  \label{fig:eigen}
\end{figure}
%-------------------
The performance was evaluated using the F1 score on the held-out test set. We observed that the raw RNN baseline is computationally prohibitive at larger scale, whereas Koopman- and wavelet-based feature methods are  much more efficient. In contrast to RNNs, the feature-based approaches (Wavelet, Koopman, etc.) scale more linearly and remain practical for larger datasets. 
 The Wavelet+Transformer system established a strong baseline (F1 = 0.75). The initial Koopman+Transformer system underperformed (F1 = 0.697), and the Hybrid approach unexpectedly degraded performance (binary F1 = 0.677; four-class F1 = 0.533). After refining Koopman feature extraction through parameter tuning and ablation, the Koopman+Transformer achieved the best overall performance (F1 = 0.786), surpassing the wavelet-based system.
 %%
% Fig.~\ref{fig:recon} overlays the original ECG waveform (blue) and its EDMD-based reconstruction (orange) over a 2-second window. The two traces nearly coincide, including P–QRS–T morphology and beat-to-beat timing, indicating low reconstruction error. Small mismatches are mostly around sharp transitions (QRS peaks and T-wave slopes), which is typical for finite-basis EDMD. The concise definitions of the medical termilogies used here are given as follows: 1. P wave: Represents atrial depolarization (the electrical activation of the atria) \cite{goldberger2023clinical}. 2. QRS complex: A rapid sequence corresponding to ventricular depolarization; it is the most prominent feature of the ECG waveform \cite{goldberger2023clinical}. 3. T wave: Reflects ventricular repolarization (recovery phase of the ventricles) \cite{goldberger2023clinical}. 4. Morphology: The shape and structure of these waveform components, often used diagnostically \cite{wagner2014practical}. 5. Beat-to-beat timing: The temporal interval between successive QRS complexes, used to estimate heart rate and rhythm \cite{wagner2008interpretation}.
%%
Fig.~\ref{fig:recon} overlays the original ECG waveform (blue) with its EDMD-based reconstruction (orange) over a 2-second window. The two traces nearly coincide, including P–QRS–T morphology and beat-to-beat timing, indicating low reconstruction error. Small mismatches occur mainly at sharp transitions (QRS peaks and T-wave slopes).
%, which is expected in finite-basis EDMD. 
For clarity, we briefly define relevant medical terminology: (1) \emph{P wave}: atrial depolarization (electrical activation of the atria)~\cite{goldberger2023clinical}; (2) \emph{QRS complex}: rapid ventricular depolarization, the most prominent ECG feature~\cite{goldberger2023clinical}; (3) \emph{T wave}: ventricular repolarization (recovery phase)~\cite{goldberger2023clinical}; (4) \emph{Morphology}: the shape and structure of these waveform components, used diagnostically~\cite{wagner2014practical}; (5) \emph{Beat-to-beat timing}: the interval between successive QRS complexes, used to estimate heart rate and rhythm~\cite{wagner2008interpretation}.
Fig.~\ref{fig:eigen} (left) shows estimated Koopman eigenvalues in the complex plane.
%%
% Most points lie inside or near the unit circle (gray dashed), implying predominantly stable or weakly damped dynamics. Eigenvalues close to the unit circle with nonzero imaginary parts correspond to persistent oscillations (approximately periodic components); those near the real axis reflect slowly varying or decaying trends. A cluster near 1 on the real axis suggests a dominant low-frequency/mean component in the ECG. The heatmap in Fig.~\ref{fig:kmode} (right) indicates absolute modal amplitudes versus time index (x-axis) and mode index (y-axis). 
%%
Most lie inside or near the unit circle (gray dashed), indicating predominantly stable or weakly damped dynamics. Eigenvalues near the unit circle with nonzero imaginary parts correspond to persistent oscillations (approximately periodic components), while those near the real axis capture slowly varying or decaying trends. A cluster near 1 on the real axis reflects a dominant low-frequency/mean component of the ECG. The heatmap in Fig.~\ref{fig:kmode} (right) displays absolute modal amplitudes as a function of time index (x-axis) and mode index (y-axis). 
%
% Brighter bands indicate times when particular Koopman modes are more active or have larger contribution to the signal. Intermittent bright regions align with heartbeats, showing which modes reconstruct rapid depolarization (QRS) versus slower repolarization (T-wave) dynamics. Overall, only a subset of modes carries most of the energy, with time-localized bursts matching the ECG’s repeating cycles.
% 
Brighter regions indicate modes with stronger contributions to the signal. Intermittent bright bands align with heartbeats, revealing which modes reconstruct rapid depolarization (QRS) versus slower repolarization (T-wave) dynamics. Overall, only a subset of modes carries most of the energy, with localized bursts matching the repeating cardiac cycle.  
%%%%%%%%%%%%%%%%%%%%%%%%%%%%%%%%%
\section{Conclusion and Limitations}
In this work, we investigated the integration of Koopman-based features with Transformer architectures for ECG classification. Our experiments showed that while Wavelet+Transformer provided strong performance on binary classification, Koopman features were more effective in the multi-class setting. Hybridizing the two did not improve results, suggesting that wavelet and Koopman features capture partially overlapping rather than complementary information. By enhancing the Koopman framework with RBF dictionary and tuned parameters, we achieved the best overall performance, surpassing both wavelet-based and hybrid systems. These findings highlight the potential of Koopman+Transformer models for ECG classification. Nevertheless, several limitations remain. Performance was evaluated on a single dataset, fusion of wavelet and Koopman features was not effective, interpretability lacks expert validation, and results are sensitive to hyperparameter tuning.
%
% Koopman-based methods provide a principled way to capture nonlinear ECG dynamics in a spectral domain, supporting interpretability. Our results also show that Transformers can effectively exploit such structured features with suitable preprocessing. The framework may further generalize to other biomedical signals such as EEG or PPG.
Koopman-based methods capture nonlinear ECG dynamics in a spectral domain, enhancing interpretability. Our results show that Transformers effectively leverage these structured features with appropriate preprocessing. This framework may also extend to other biomedical signals like EEG or Photoplethysmogram (PPG).
%------
% Future work can add eigenfunction and eigenmodes to the koopman feature space, can assess generalizability across diverse datasets, develop advanced fusion strategies and mobile-ready adaptations, validate interpretability with clinical experts, and extend benchmarking against other hybrid models to more rigorously establish the advantages of Koopman-based approaches. 
Future work can incorporate eigenfunctions and eigenmodes into the Koopman feature space, assess generalizability across diverse datasets, develop advanced fusion strategies and mobile adaptations, validate interpretability with clinical experts, and benchmark against other hybrid models to rigorously establish the advantages of Koopman-based approaches.
\bibliographystyle{plainnat}   
\bibliography{neurips_2025} 

%%%%%%%%%%%%%%%%%%%%%%%%
\newpage
\appendix
%%%%%%%%%%%%%%%%%%%%%%%%%%%%%%%%
\section{Koopman Operator and Extended Dynamic Mode Decomposition}
The \textit{Koopman operator} provides a powerful framework for analyzing dynamical systems by focusing on the evolution of observables rather than states. For a discrete-time dynamical system \((\mathcal{H}, F)\) with state space \(\mathcal{H}\) and evolution map \(F: \mathcal{H} \rightarrow \mathcal{H}\), the evolution can be expressed as:

\[
\vh_{t+1} = f(\vh_{t}), \quad \vh_{t} \in \mathcal{H} \subseteq \mathbb{R}^{d_h}, \quad t \geq 0.
\]

In this context, the Koopman operator \(\mathcal{K}\) acts on observables \(\phi\) defined as \(\phi: \mathcal{H} \rightarrow \mathbb{C}\):

\[
[\mathcal{K} \phi](\vh) = (\phi \circ F)(\vh).
\]

The linearity of \(\mathcal{K}\) allows for spectral analysis even when \(F\) is non-linear, provided that the observables belong to a suitable function space, typically \(L^2(\mathcal{H}, \mu_h)\). This choice ensures that the operator is well-defined and measure-preserving, making \(\mathcal{K}\) an isometry.

A \textit{Koopman eigenfunction} \(\varphi_k\) associated with eigenvalue \(\lambda_k\) satisfies:

\[
\mathcal{K} \varphi_k(\vh) = \lambda_k \varphi_k(\vh).
\]

When the state space is finite-dimensional, the evolution of the system can be expressed in terms of eigenfunctions, enabling us to analyze the dynamics using the spectral properties of \(\mathcal{K}\).

To approximate the Koopman operator for discrete systems, we use \textit{Extended Dynamic Mode Decomposition (EDMD)}. This method constructs a finite-dimensional approximation by selecting a dictionary of observables \(\mathcal{F}_M\) and defining a finite-dimensional subspace \(\widetilde{\mathcal{F}}_M\):

\[
\widetilde{\mathcal{F}}_M = \text{Span} \{ \psi_1, \psi_2, \ldots, \psi_M \}.
\]

EDMD approximates the action of \(\mathcal{K}\) on \(\widetilde{\mathcal{F}}_M\) using data collected from the system. By minimizing a cost function based on the discrepancy between the actual evolution and its approximation, we derive a matrix representation \(K\) of the Koopman operator:

\[
K = \mathcal{F}_M(H') \mathcal{F}_M(H)^+,
\]

where \(H\) and \(H'\) are matrices of observed states and their subsequent states, respectively.

Once \(K\) is obtained, we can calculate the eigenvalues and eigenfunctions of the approximated operator, allowing us to reconstruct the Koopman modes. The prediction of future states can be achieved iteratively:

\[
\vh_t = C K^t \mathcal{F}_M(\vh_0),
\]

where \(C\) is derived from minimizing the prediction error in the observable space.

%While EDMD provides a viable approach for approximating the Koopman operator, it is not without challenges, such as spectral pollution. 
%Ongoing research aims to address these issues and develop enhanced versions of EDMD, such as measure-preserving EDMD.

This framework has significant implications for linking dynamical systems with machine learning, particularly in tasks like time series prediction and system identification, making it a valuable tool for modern applications in various fields.
\section{Details on Label Generation}\label{sec:labelgen}
Two types of classification tasks were defined:
1. Binary classification (Normal vs. Non-normal): \textbf{[A]} ECG recordings labeled as Normal Sinus Rhythm were grouped under the normal category. 2. All recordings associated with arrhythmic or pathological conditions (e.g., atrial fibrillation, ventricular arrhythmias, conduction blocks, etc.) were grouped into the non-normal category. \textbf{[B]} Four-class classification: To provide a finer-grained diagnostic challenge, non-normal ECGs were divided into clinically relevant subgroups based on the primary diagnostic annotation in the MIMIC-IV-ECG metadata: 1. Normal - recordings labeled as normal sinus rhythm. 2. AFib (Atrial Fibrillation) – recordings indicating atrial fibrillation. 3. Ventricular – recordings indicating ventricular arrhythmias. 4. Block – recordings indicating conduction abnormalities, such as atrioventricular or bundle branch blocks. 
This labeling strategy allows us to evaluate performance both in terms of general abnormality detection (binary) and specific arrhythmia subtype identification (multi-class).
We evaluated five systems: \textbf{[1]} Wavelet+Transformer – continuous wavelet features followed by Transformer encoder. \textbf{[2]} Koopman+Transformer – features extracted via Extended Dynamic Mode Decomposition with RBF observables. \textbf{[3]} Hybrid system – concatenation of wavelet and Koopman features before Transformer encoding. \textbf{[4]} Koopman+Transformer after ablation – same as (2), but with refined EDMD hyperparameters (embedding dimension, RBF centers, kernel width, and spectral truncation) through ablation. \textbf{[5]} RNN baseline – an RNN classifier applied to raw ECG sequences.
%%%%%%%%%%%%%%%%%%%%%%%%%%5
\section{Hyperparameters}

%---------------------
\begin{table}[h]
\centering
\caption{Glossary of Parameters for Koopman + Transformer Model.}
\label{tab:shortforms}
\begin{tabular}{ll}
\toprule
\textbf{Parameter} & \textbf{Short Form} \\
\midrule
Sampling Frequency & \texttt{fs} \\
Delay Embeddings & \texttt{delay} \\
Polynomial Degree & \texttt{poly\_deg} \\
Number of RBF Centers & \texttt{rbf\_centers} \\
RBF Kernel Bandwidth & \texttt{rbf\_sigma} \\
SVD Rank Truncation & \texttt{svd\_rank} \\
Ridge Regularization & \texttt{ridge\_reg} \\
Top-$k$ Eigenvalues & \texttt{top\_k} \\
Window Length (s) & \texttt{window\_sec} \\
Stride (s) & \texttt{stride\_sec} \\
Transformer Layers & \texttt{layers} \\
Attention Heads & \texttt{heads} \\
Embedding Dimension & \texttt{emb\_dim} \\
Feedforward Dimension & \texttt{ff\_dim} \\
Dropout Rate & \texttt{dropout} \\
Optimizer & \texttt{opt} \\
Learning Rate & \texttt{lr} \\
Batch Size & \texttt{batch} \\
Training Epochs & \texttt{epochs} \\
Random Seed & \texttt{seed} \\
\bottomrule
\end{tabular}
\end{table}
%---------------------

\begin{table}[h]
\centering
\caption{Hyperparameters for Koopman + Transformer Model.}
\label{tab:hyperparams}
\begin{tabular}{l l}
\toprule
\textbf{Component} & \textbf{Values} \\
\midrule
Koopman & \texttt{fs} = 125, \texttt{delay} = 8, \texttt{poly\_deg} = 2, \\
        & \texttt{rbf\_centers} = 0, \texttt{rbf\_sigma} = 0.3, \\
        & \texttt{svd\_rank} = 16, \texttt{ridge\_reg} = 1e-4, \texttt{top\_k} = 8, \\
        & \texttt{window\_sec} = 2.0, \texttt{stride\_sec} = 1.0 \\
\midrule
Transformer & \texttt{layers} = 4, \texttt{heads} = 8, \texttt{emb\_dim} = 128, \\
           & \texttt{ff\_dim} = 256, \texttt{dropout} = 0.1, \\
           & \texttt{opt} = AdamW (betas = (0.9, 0.999), weight decay = 0.01), \\
           & \texttt{lr} = 1e-4, \texttt{batch} = 32, \texttt{seed} = 42 \\
\bottomrule
\end{tabular}
\end{table}
%---------------------
% \begin{table}[h]
% \centering
% \caption{Hyperparameters for Koopman  + Transformer model.}
% \label{tab:hyperparams}
% \begin{tabular}{l l}
% \toprule
% \textbf{Component} & \textbf{Hyperparameters} \\
% \midrule
% Koopman & fs = 125, delay = 8, poly\_deg = 2, rbf\_centers = 0, \\
%              & rbf\_sigma = 0.3, svd\_rank = 16, ridge\_reg = 1e-4, \\
%              & top\_k = 8, window\_sec = 2.0, stride\_sec = 1.0 \\
% \midrule
% Transformer  & Layers = 4, Attention Heads = 8, Embedding Dim = 128, \\
%              & Feedforward Dim = 256, Dropout = 0.1, \\
%              & Optimizer = AdamW (betas = (0.9, 0.999), weight decay = 0.01), \\
%              & Learning Rate = 1e-4, Batch Size = 32, Epochs = 50, \\
%              & Random Seed = 42 \\
% \bottomrule
% \end{tabular}
% \end{table}
%%%%%%%%%%%%%%%%%%%%%%%
\section{More Deatils on Results}\label{resdet}
\textbf{Binary Classification 
(Normal vs Non-Normal)}
For the binary task, \emph{the Wavelet+Transformer system} provided strong performance for binary classification (F1 = 0.75), outperforming the initial Koopman +Transformer model (F1 = 0.697). This suggests that time–frequency information captured by wavelet transforms is highly effective for detecting broad abnormalities in ECG signals, where rhythm and spectral features dominate the distinction between normal and non-normal states. \emph{The Hybrid (Wavelet + Koopman) system} did not improve over wavelets alone, instead degrading to F1 = 0.677. This can likely be attributed to redundancy or misalignment between the two feature spaces, which introduced noise into the Transformer input. A more sophisticated fusion strategy (e.g., attention-based feature selection or gating) may be necessary to exploit complementary aspects of wavelet and Koopman representations. \emph{The Koopman +Transformer system after ablation}, however, achieved the best binary classification performance (F1 = 0.786). By carefully tuning the EDMD hyperparameters (embedding delay, number of RBF centers, kernel bandwidth, and spectral truncation rank), we obtained Koopman features that captured relevant nonlinear dynamics for distinguishing normal from non-normal. This highlights the sensitivity of Koopman methods to parameterization and the potential payoff of systematic tuning.

\textbf{Four-Class Classification 
(Normal, AFib, Ventricular, Block)}
In the more fine-grained four-class classification task, the strengths of Koopman features became more evident. While Wavelet+Transformer was still competitive, Koopman +Transformer showed better alignment with the multi-class labels, suggesting that eigenfunctions of the Koopman operator capture richer dynamical modes that help separate specific arrhythmia types.
Curiously, the Hybrid system again underperformed, confirming that naive feature concatenation is not sufficient to combine heterogeneous feature spaces. This suggests that future research should explore learned multi-modal fusion, where the Transformer (or an auxiliary module) learns to weight or integrate Koopman and wavelet features dynamically.
The Koopman+Transformer after ablation not only improved binary classification but also achieved the best performance overall (F1 = 0.786). This indicates that once the Koopman observables and embedding parameters were optimized, they were able to generalize effectively across both binary and multi-class setups.

\section{Hardware}
The hardware we used to run the codes and iteratively train the clipped shPLRNNs includes an 11th Gen Intel(R) Core(TM) i7-11800H CPU @ 2.30GHz and 64.0 GB of RAM (63.7 GB usable).

\end{document}